  \documentclass{aa}
 \usepackage{graphicx}
 \newcommand{\Mv}{$\mathrm{M_{v}}$}
 \newcommand{\Mbol}{$\mathrm{M_{bol}}$}
 \newcommand{\R}{log$R^{'}_{\rm HK}$}
 \newcommand{\Rx}{$R_{\rm x}$}
 \newcommand{\Vt}{$\mathrm{V_{t}}$}
 \newcommand{\vsini}{$v$sin$i$}
 \newcommand{\Teff}{$T_{\rm eff}$}
 \newcommand{\logg}{$\log \mathrm{g}$}
 \newcommand{\FeH}{$\mathrm{[Fe/H]}$}
 \usepackage{aalongtable}
\begin{document}
 \title{Activity and the Li abundances in the FGK dwarfs
 \thanks{Based on the spectra collected with the ELODIE spectrograph using the
 1.93-m telescope at the Observatoire de Haute Provence (CNRS, France).}
 \thanks{Table 3 is only available in electronic form at the CDS
 via the anonymous ftp to cdsarc.u-strasbg.fr (130.79.128.5)
 or via http://cdsweb.u-strasbg.fr/cgi-bin/qcat?J/A+A/(vol)/(page)}}
 \author{T.~V.~Mishenina,
 \inst{1,2}\,\
 C.~Soubiran,
 \inst{2}\,\
 V.~V.~Kovtyukh,
 \inst{1}\,\
 M.~M.~Katsova,
 \inst{3}\,\
  and M.~A.~Livshits
 \inst{4}}
 \offprints{T.~V.~Mishenina}
 \institute{Astronomical Observatory, Odessa National University,
 and Isaac Newton Institute of Chile, Odessa Branch,
 T.G. Shevchenko Park, 65014, Odessa, Ukraine\\
 \email {tamar@deneb1.odessa.ua}
 \and
 Universit\'e Bordeaux 1 -- CNRS -- Laboratoire d'Astrophysique
  de Bordeaux, UMR 5804, 33271 Floirac Cedex, France\\
 \email {Caroline.Soubiran@obs.u-bordeaux1.fr}
 \and
  Sternberg State Astronomical Institute, Lomonosov Moscow State University,
  13, Universitetsky av., 119991 Moscow, Russia\\
  \email {maria@sai.msu.ru}
 \and
  Pushkov Institute of Terrestrial Magnetism, Ionosphere, and Radio Wave Propagation of
 Russian Academy of Sciences (IZMIRAN), Troitsk, 142190,
  Moscow, Russia
  \email {maliv@mail.ru}
 }
 \date{Received \today; accepted }
 \authorrunning{Mishenina et al.}
 \titlerunning{Li abundance...}

 \abstract
   {}
   {The aim of the present study is to determine the Li abundances for a large set of the FGK
   dwarfs and to analyse the connections between the Li content, stellar parameters, and activity.}
     {The atmospheric parameters, rotational velocities and the Li abundances were determined from
   a homogeneous collection of the echelle spectra with high resolution and a high signal-to-noise ratio.
   The rotational velocities \vsini\ were determined by calibrating  the cross-correlation function.
   The effective temperatures \Teff\ were estimated by the line-depth ratio method. The surface
   gravities \logg\ were computed by two methods: the iron ionization balance and the parallax.
    The LTE Li  abundances were computed using synthetic spectra method.
   The behaviour of the Li abundance was examined in correlation
   with \Teff, \FeH,  as well as with \vsini\ and the level of activity in three   stellar groups of the different  temperature range.}
   {The stellar parameters and the Li abundances are presented for 150 slow rotating stars of the lower part of MS.
     The studied stars show a decline in
   the Li abundance with decreasing temperature \Teff\ and a significant spread,
   which should be due to the difference of age of stars.
   The correlations between the Li abundances, rotational velocities
   \vsini, and the level of
   the chromospheric activity were discovered for the stars with 6000$>$\Teff$>$5700 K,
   and it is tighter for the stars with 5700$>$\Teff$>$5200 K.
  The target stars  with  \Teff$<$5200 K do not show any correlation between log A(Li) and \vsini.
  The relationship between the chromospheric and coronal fluxes in active with detected Li as well as in less
  active stars gives a hint that there exist different conditions in the action of the dynamo mechanism in those stars.}
   {We found that the Li-activity correlation  is evident only
   in a restricted temperature range and
   the Li abundance spread seems to be present in a group of low chromospheric
   activity stars  that  also show  a broad spread in  the
   chromospheric vs. coronal activity.
   }

   \maketitle

 \section{Introduction}

  Li burns at temperatures of $2.5\cdot10^6$ K via $\alpha$
 captures, which makes it a useful probe of the mixing in stars.
 The connection of the Li abundance with stellar activity has a long
 history (Herbig \cite{herb65}, Pallavicini et al. \cite{pal87}). It was soon realized
 that, like stellar activity, high Li abundance is found in
 young stars, although the Li abundance, by itself, is not sufficient
 to estimate the age of a star. Skumanich (\cite{sku72}) showed that both the
 chromospheric emission of the Ca II H and K lines and the rotational
 velocity decline as the square root of the age of stars ({\it the Skumanich law}).
 However, the Li abundance does not show a tight correlation
 with either (Duncan \cite{dun81}, Duncan \& Jones \cite{dun83},
  Soderblom et al. \cite{sod93a}).
 Whether  a tight relation exists  between the Li abundance and
 the stellar mass, the spread, observed in the Li abundance owing to
 presence of spots on the star's surface (Soderblom et al. \cite{sod93a}) or
 to the main-sequence (MS) depletion (Thorburn et al. \cite{thor93})
is still an open  question.

 A spread of the Li abundance in the open cluster (OC) stars and binaries
 with the same mass and the Li depletion (Martin et al. \cite{mart02},
 King \cite{king10}), as well as higher Li abundance of
  rapidly rotating dwarfs (Cutispoto et al. \cite{cut03}), are not corroborated
 by the theoretical predictions of the near-solar masses, in which the
 convection was only considered  as
 a mixing mechanism   (Randich \cite{ran10}),
 and fast rotation intensifies the mixing process that leads to the Li destruction
 (e.g. Charbonnel et al. \cite{char92}).

 The observation of the rotation and chromospheric emission in the F, G, and K dwarfs
 of the OC stars allowed  a reliable
 age-rotation relationship to be established that is one of the bases of the method
 of  estimating  age depending on
 the level of activity, so-called gyrochronology
 (Soderblom et al. 1993bc, Barnes \cite{barn03}).
 That paradigm can serve as a
 basis for understanding the relationship of activity, the light elements
 content and rotation (Barnes \cite{barn03}).

 Analysis of the mechanisms that provided the  Li overabundance
 was carried out for the stars of the RS CVn -- type binaries with spots and
 a significant rate of rotation (Pallavicini et al. \cite{pal92}, Pallavicini et al. \cite{pal93}, Randich et al. \cite{ran93}).
 In  doing so, the moderate excess (overabundance) of  Li compared to normal stars of the same spectral types was confirmed, and the correlation with the rotation parameter
 and the chromospheric fluxes was not discovered. The possibility of the Li enrichment,  connected to the presence of spots and the production of
 additional Li in areas (Pallavicini et al. \cite{pal92}), was not eliminated.
  The fresh isotopes
 of Li  can be produced  by the nuclear interactions of ions,
 accelerated at the surface of the flaring stars (see e.g. Canal et al. \cite{Can75},
 Livshits \cite{Liv97}, Tatischeff \& Thibaud \cite{tat07}). Only one observational result that
 attests to the Li growth in the stellar area was obtained by Montes \& Ramsey (\cite{mon98}).

   The chromospheric and coronal activities of  late-type stars have been
studied more in  past decades (Guedel \cite{Gud04}, Guinnan \& Enge \cite{GuiEn09},
Katsova \& Livshits \cite{katliv11}).
 Katsova \& Livshits (\cite{katliv11}) find that the ideas about
 the gyrochronology (Mamajek \& Hillenbrand \cite{mam08})
 is valid for the stars that are hotter than the Sun (with \Teff$>$ \Teff $_{\sun}$),
but for  the late-type dwarfs  stars (with \Teff $<$ \Teff $_{\sun}$), whose convection zones
 are thicker than the solar ones, the activity  evolves in time apparently according to  the other  law.
Studing those stars that are  both hotter and cooler
   than the Sun,  can clarify the understanding of the features and causes of activity of
   the FGK dwarfs, as well as the real relation between the enhancement of Li and activity.

   In this study we aim  at finding some observational evidence of  the Li abundance - activity behaviour
   in the stars in the lower part of the MS,
   that are the slow rotating dwarfs with masses close to the solar one.

  The paper is divided as follows. The observational data are described in Sect. 2;
 Sect. 3 is devoted to the description of the levels of activity of the investigated stars;
 the methods and errors in determining of the rotational velocities, atmospheric parameters and
 the Li abundances are presented in Sections 4, 5, and 6, respectively. In Sect. 7,
 the Li
 abundance behaviour is considered  in relation with the stellar and activity parameters.
 Conclusions are drawn in Sect.8.

 \section{Observations and the spectral processing}

 The spectra of 150 stars were obtained using the 1.93\,m telescope at the Observatoire de
 Haute-Provence (OHP, France), equipped with the \'echelle type spectrograph ELODIE
 (Barrane et al. \cite{ba96}) which provides a resolving power of R = 42\,000. Most of the target
 stars,   126 stars on the lower part of the MS that are examined in the present study,
 were taken from our previous paper Mishenina et al. (2008) (M08). Some stars from M08 were
 eliminated from the analysis because it was impossible to estimate the Li abundances owing to distortions
 of the spectra in the Li line region. A set of 24 stars was added with spectra that were  previously analysed
 by Mishenina et al. (\cite{mish04}) or retrieved from the ELODIE archive
 (Moultaka et al. \cite{moul04}). All 150 spectra were processed homogeneously using the same methods.

\begin{figure}
\includegraphics[width=8.8cm]{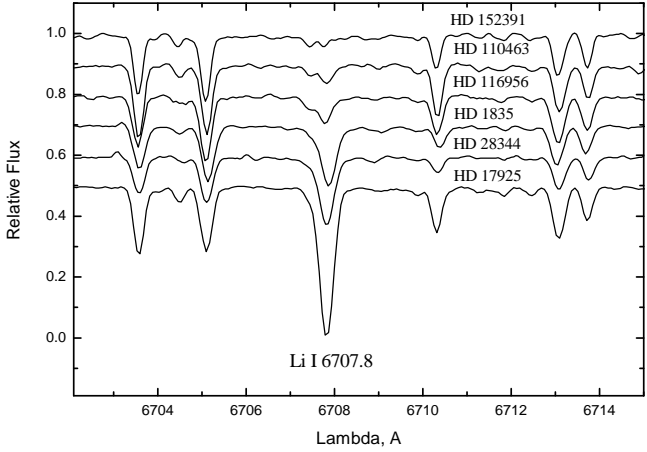}
\caption{The spectra in the Li 6707 \AA\ line region for some of the target stars.}
\label{specli}
\end{figure}

 The extraction of the 1D spectra and measurement of the radial velocities were performed with
the standard on-line ELODIE reduction software, while the deblazing and removal of cosmic particles
were carried out following Katz et al. (\cite{katz98}). Further processing of spectra
(the continuous spectrum level set up and measurement of the equivalent widths) was conducted
using the DECH20 software package (Galazutdinov \cite{gal92}).
Figure \ref{specli} shows the Li region in the spectra for some target stars.

\section{Levels of activity of the investigated stars}

In \cite{mish08}, we investigated the difference between the elemental abundances in active and
inactive stars. We labelled as active those stars  of the BY Dra type, RS CnV type, and flare Fl stars
according to  SIMBAD. We also classified six variable stars as active that showed  evidence
of the chromospheric activity in their \ion{Ca}{ii} and H$_{\rm \alpha}$ lines (see details
in \cite{mish08}). In the present study,
we  also considered  active  those
with a high level of activity, as discussed below.

 As an indicator of the level of the chromospheric activity we examined the corresponding
index \R\ that measures the chromospheric emission in the cores of the broad photospheric
 \ion{Ca}{ii} H and K absorption lines, normalized to the underlying photospheric spectrum.
For a part of the target stars with \Teff$>$ 4800 K, we found the \R\ values in
Wright et al. (\cite{wrig04}). We also used the data from the studies by
Henri et al. (\cite{henry96}), Hall et al. (\cite{hall07}), Baliunas et al. (\cite{bal95}), and
Maldonado et al.(\cite{mal10}). A total of 84 different stars were retrieved from those catalogues.
The values of \R\  by Wright et al. (\cite{wrig04}) are compared to those from different studies
(Table \ref{table:rhk}), and agreement is good.
For some  stars where the \R\ index was not detected, we determined it on the base
of the correlation between the chromospheric and the X-ray (Mamajek \& Hillenbrand \cite{mam08}).
The X-ray  was taken from  the ROSAT All-Sky Survey (Voges
et al. \cite{Voges99}).

 On having applied this index, it was necessary for us to pick out its value, corresponding to
the ``section boundary" between  active and inactive stars. It should be noted that the values of indices themselves
differ from each other in various studies. The discussion for some of these stars can be seen in the Appendix.

  We accepted that the transition between active and inactive stars occurs
at \R$\sim$--4.75. As indicated by Lovis et al. \cite{lov11} ``this value corresponds to active
stars according to the limit given by the Vaughan-Preston gap (Vaughan \& Preston \cite{vaug80}".
Moreover, a close value of \R$\sim$--4.80 was declared in the study by Jenkins et al. (\cite{jen11})
and Katsova \& Livshits (\cite{katliv11}).

%Table 1
\begin{table}
  \caption[]{Comparison of the \R\  obtained by
  Wright et al. (\cite{wrig04}) and  from other sources}
\label{table:rhk}
\begin{tabular}{rccl}
\hline
\hline
   $\Delta$\R & $\sigma$  & N & sources \\
\hline
   0.00& $\pm$0.09 &6& Henry et al. (\cite{henry96})  \\
   0.01& $\pm$0.04 &7& Baliunas et al. (\cite{bal95})\\
 --0.01& $\pm$0.07 &7& Hall et al. (\cite{hall07})    \\
\hline
\end{tabular}
\end{table}

%fig1

%Fig2
\begin{figure}
\includegraphics[width=8.8cm]{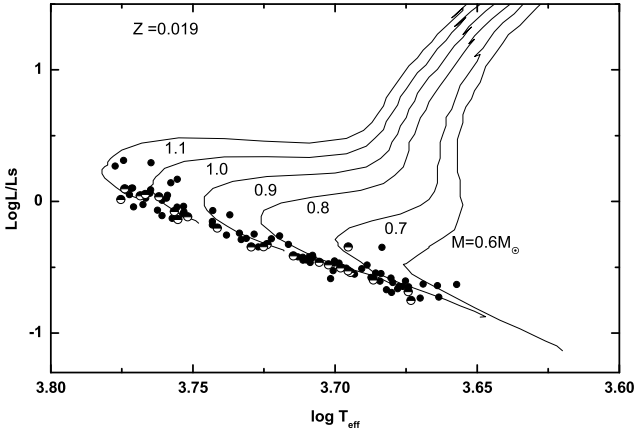}
\caption{Position of our target stars in the log L/L$_{\sun}$ vs. log \Teff\ diagram.  Evolutionary
tracks were taken from Girardi et al. \cite{gir00}, the stars of the BY Dra types are marked as
semi-full circles, the other stars  as full circles. }
\label{LLT}
\end{figure}

%fig3
\begin{figure}
\includegraphics[width=8.8cm]{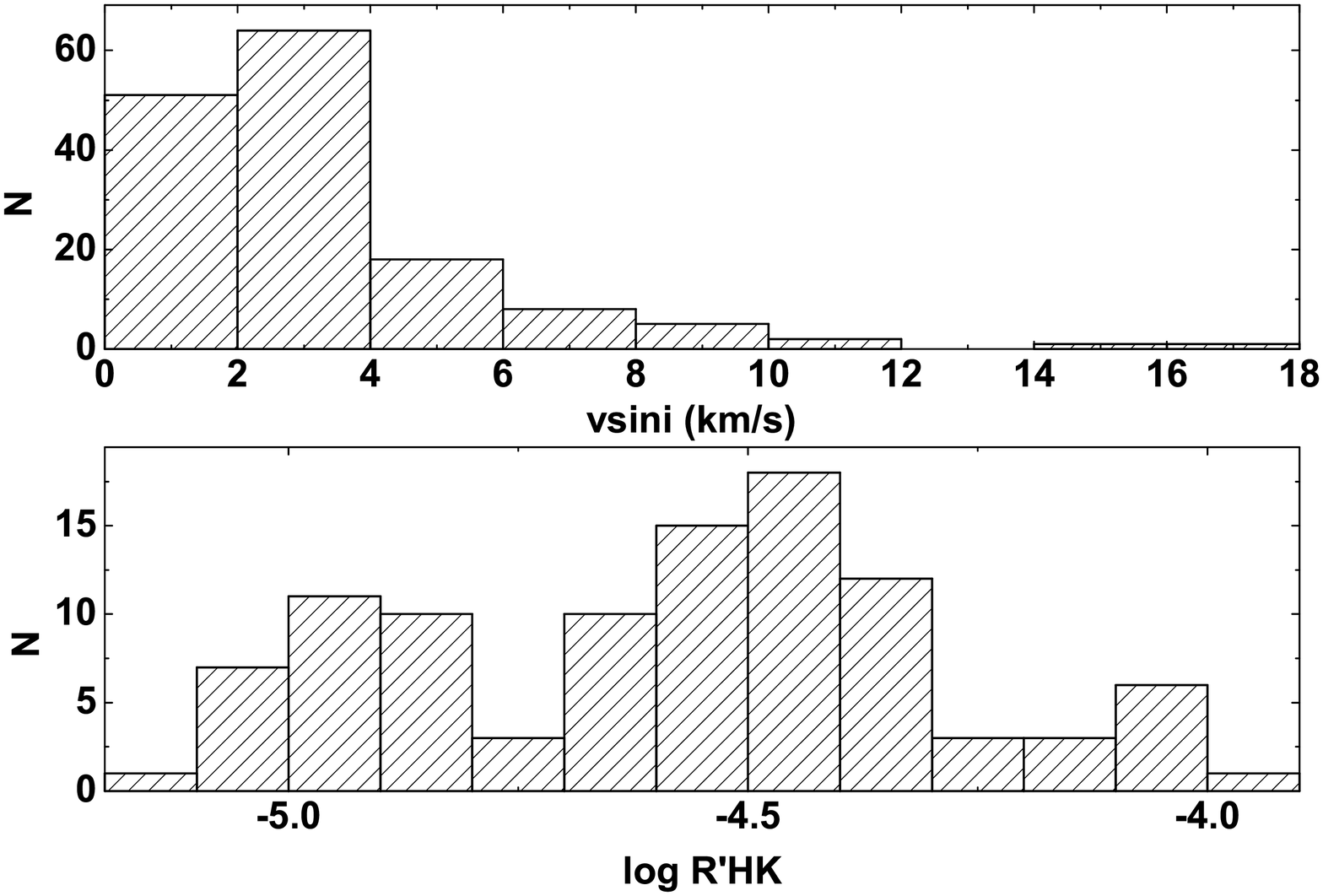}
\caption{The distribution of  \vsini\ and \R\ of our target stars}
\label{gist}
\end{figure}

  %fig4
  \begin{figure}
  \includegraphics[width=8.8cm]{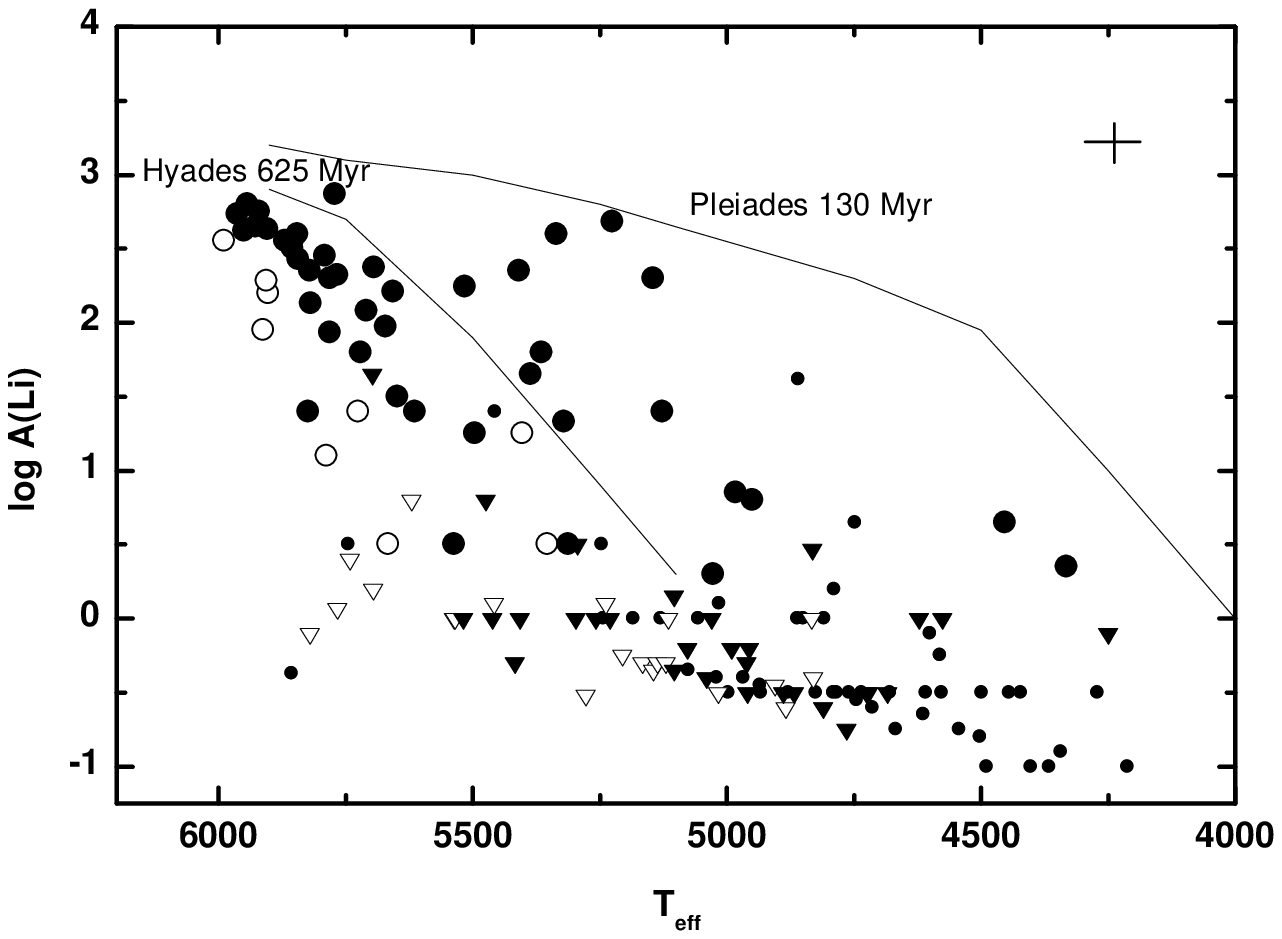}
  \caption{Lithium abundances vs. \Teff. The medium trend of the Pleiades and the Hyades is represented
according to Soderblom et al. (\cite{sod93a}) and Thorburn et al. (\cite{thor93}),
respectively. The stars with a high level of the chromospheric activity (\R $>$--4.75) are marked
as full circles, those  with a weak level of activity (\R $<$--4.75)  as open circles. Triangles
denote the upper limits of the Li determinations. The stars, for which \R\ are not defined,
are marked as small full circles.}
  \label{liTeff}
  \end{figure}

\section{Rotational velocities}

We  measured the rotational velocities (\vsini) of the target stars with a relation
calibrated by Queloz et al. (\cite{que98}), giving \vsini\ as a function of $\sigma {RV}$,
the standard deviation of the ELODIE cross-correlation function, approximated by the Gaussian.
The relation is the following:
$$ vsini =1.90\sqrt{\sigma_{RV}^2-\sigma_0^2}.$$

The parameter $\sigma_0$ represents the mean intrinsic width for the non-rotating stars. It was
calibrated by Queloz et al. (\cite{que98}) as a function of (B--V): $$\sigma_0 = 0.27(B-V)^2+4.51.$$
This \vsini\ calibration is valid for the stars in the colour range $ 0.7 \leq B-V \leq 1.4$, with
slow and moderate rotation rates ($\leq 40\, \rm{km s}^{-1}$) and the solar metallicities that are
consistent with our stars.

 We compared our values of \vsini\ with the results of determinations by other authors
(Table \ref{table:tabcompvs}).
The comparison of our \vsini\ determinations with six other studies shows good compliance.
The low values of \vsini\ are the upper limits of the projection of the rotational velocity
definition. The obtained values of \vsini\ are given in Table \ref{table:tabsumm}.

 %Table2
\begin{table}
\caption[]{Comparison of our $v\sin i$ values with those of other authors
(the mean difference and standard deviation are presented).}
\label{table:tabcompvs}
%\tiny
\begin{tabular}{rccl}
\hline
\hline
$\Delta$$v\sin i$ & $\sigma$ & n & source\\
\hline
--0.15& $\pm$1.33 & 83 & Nordstr\"om et al. (\cite{nor04}) \\
--0.37& $\pm$1.18 &  67 &  Valenti \& Fisher (\cite{val05})\\
--0.70& $\pm$2.00 & 26 & Tokovinin (\cite{tok92})\\
--0.29& $\pm$0.71 & 25 & Fuhrmann (\cite{fuhr08})\\
  0.13& $\pm$0.96 & 14 & Benz \& Mayor (\cite{ben84})\\
--0.56& $\pm$0.67 & 12 & Gaidos et al. (\cite{ga00})\\
\hline
\end{tabular}
\end{table}

\section{Atmospheric parameters}

 We defined parameters (\Teff, \logg, \FeH)  just for several newly
included stars and for the others we used the determinations by Mishenina et al. (\cite{mish04})
and M08. The effective temperatures \Teff\ were estimated by the line-depth ratio method (Kovtyukh et al. \cite{kovt04}).
The surface gravity \logg\ was computed by two methods: the iron ionization balance \logg$_{IE}$
and the parallax \logg$_P$.
The results of applying of those two methods are in good compliance (\cite{mish08}).
 For cooler stars, which are short in the  \ion{Fe}{ii} lines, the parallax was the only method used.
The microturbulent velocity \Vt\ was derived by considering that the iron abundance log A(Fe),
obtained from the given \ion{Fe}{i} line, is not correlated with the equivalent width (EW) of
that line.
The adopted metallicity [Fe/H] is the iron abundance, which was determined from the \ion{Fe}{i} lines.

 The comparison of the obtained atmospheric parameters with those by the other authors is
given in Table \ref{table:tabcomp}. As is evident, there is no significant difference
between various determinations. The errors of the obtained temperature determinations are
given for each star, once determined by the line-depth ratio method. For \logg, \Vt, and \FeH,
we estimated the errors to be 0.2 dex, 0.2 dex, and 0.05 dex, respectively.

 %Table 3
\setcounter{table}{2}
\begin{table*}
\begin{minipage}{14cm}
\caption[]{The stellar parameters, determined in the present study, \logg (1): spectroscopic,
\logg(2): astrometric,  object type from SIMBAD, Rx adopted from the ROSAT catalogues or estimated
from the ROSAT archive data, indices  \R retrieved from different sources, the measured Li abundance
and the class of activity.}
 \label{table:tabsumm}
\begin{tabular}{rrrrrrrrrlrr}
\hline
\hline
    HD       & \Teff & \logg (1) & \logg (2) & \FeH  & vsini & object type & Rx & \R & Ref  &  log A(Li)  & class  \\
\hline
     166     & 5514  & 4.6   &  4.6  & 0.16  &  4.0  &   BY  &--4.31 &--4.33 &               &  2.24      & A      \\
    1835     & 5790  & 4.5   &  ...  & 0.13  &  6.51 &   BY  &--4.58 &--4.44 & W04           &  2.45      & A      \\
    3651     & 5277  & 4.5   &  4.5  & 0.15  &  0.0  &   V   &--6.07 &--5.02 & W04           &  $<$--0.52 & wA      \\
    4256     & 5020  & 4.3   &  4.3  & 0.08  &  1.5  &   ... &  ...  &  ...  &               &  $<$--0.40 & ...   \\
    4628     & 4905  & 4.6   &  ...  &--0.21 &  1.5  &   ... &--5.93 &--4.85 & H96, B95      &  $<$--0.45 & wA     \\
    4635     & 5103  & 4.4   &  4.4  & 0.07  &  1.4  &   ... &  ...  &--4.67 & W04           &  $<$0.15   & A     \\
    4913     & 4342  & 4.4   &  ...  & 0.05  &  2.15 &   ... &  ...  &  ...  &               &  $<$--0.90 & ...   \\
    6660     & 4759  & 4.6   &  ...  & 0.08  &  2.5  &   ... &  ...  &  ...  &               &  $<$--0.50 & ...   \\
    7590     & 5962  & 4.4   &  4.4  &--0.10 &  6.7  &   BY  &--4.69 &--4.53 & W04           &  2.73      & A      \\
    7924     & 5165  & 4.4   &  4.4  &--0.22 &  1.1  &   ... &--5.71 &--4.83 & W04           &  $<$--0.30 & wA     \\
    ...      & ...   & ...   &  ...  & ...   &   ... &   ... &  ...  &  ...  &  ...          &   ...      & ...    \\
\hline
\end{tabular}
\end{minipage}
\\
Notation. The \R\ values, taken from Wright et al. (\cite{wrig04}) are marked by W04;
from Henri et al. \cite{henry96} - marked by H96; from Hall et al. (\cite{hall07}) - by H07;
from Baliunas et al. (\cite{bal95}) - by B95, from  Maldonado et al.(\cite{mal10}) - by M10.
The stars with weak levels of the solar-type activity are
marked by wA; the stars with high levels of the solar-type activity are marked by A in
the last column.
\end{table*}

 %Table 4
\begin{table*}
\caption[]{The comparison  of  the atmospheric parameters obtained by us with
those by the other authors.}
\label{table:tabcomp}
\tiny
\begin{tabular}{rcccl}
\hline
\hline
$\Delta$T{$_{eff}$} & $\Delta$$\log g$        & $\Delta$[Fe/H]  & n & source\\
\hline
 16 $\pm$ 68 &                 &                 & 32 & Masana et al. (\cite{mas06}) \\
--8 $\pm$ 63 & --0.14$\pm$0.18 &--0.03 $\pm$ 0.08& 67 & Valenti \& Fisher (\cite{val05}) \\
 42 $\pm$ 36 & --0.07$\pm$0.15 &  0.00 $\pm$ 0.05& 26 & Fuhrmann (\cite{fuhr08}) \\
 1  $\pm$ 73 & --0.01$\pm$0.17 &--0.03 $\pm$ 0.04& 14 & Gaidos \& Gonzalez (\cite{gai02})\\
\hline
\end{tabular}
\end{table*}

%Table 5
\begin{table*}
\caption[]{The influence of the stellar parameters on the Li abundance
determination.}
\label{table:3stars}
\begin{tabular}{ccccccc}
\hline
\hline
\multicolumn{5}{c}{Parameter deviations}&\multicolumn{2}{c}{Total error}\\
\hline
\hline
$\Delta$\Teff & $\Delta$\logg & $\Delta$\Vt &  EW=100 m\AA &  EW = 5m\AA\  &strong lines  &weak lines  \\
 --100 K      &  --0.2        & +0.2        &   +3 m\AA            &   +2 m\AA             &       &    \\
\hline
\multicolumn{6}{c}{ $T_{eff}$=5800, $\log~g $=4.5, Vt =1.0 km/s, [Fe/H]=0.0}   \\
   0.095 &     0.001  &  0.009    &    0.03   &          0.15   &          0.10  &   0.18  \\
\hline
\multicolumn{6}{c}{$T_{eff}$=5000, $\log~g $=4.5, Vt =1.0 km/s, [Fe/H]= 0.0}\\
   0.113  &    0.003  &  0.006  &      0.03   &           0.15 &           0.12  &  0.19 \\
\hline
\multicolumn{6}{c}{ $T_{eff}$=4600, $\log~g $=4.5, Vt =1.0 km/s, [Fe/H]= 0.0}\\
 0.141&    0.000  &   0.008 &      0.03  &            0.15 &              0.14 &  0.20 \\
\hline
\hline
\end{tabular}
\end{table*}

%Table 6
\begin{table*}
\caption[]{The comparison of the atmospheric parameters and the Li abundance
with those obtained by the other authors.}
\label{table:tabcomp2}
\tiny
\begin{tabular}{lccccl}
\hline
\hline
$\Delta$\Teff & $\Delta$\logg & $\Delta$[Fe/H]& A(Li) & n & source\\
\hline
67 $\pm$ 137 &  0.23$\pm$0.11& 0.11 $\pm$0.08&--0.01$\pm$0.10 & 3& Chen et al. (\cite{chen01}) \\
76 $\pm$  69 &               &               &  0.09$\pm$0.14 & 3& Favata et al. (\cite{fav97}) \\
29 $\pm$  67 &  0.06$\pm$0.17& 0.03 $\pm$0.06&--0.08$\pm$0.12 &10& Takeda et al. (\cite{tak05}, \cite{tak07}) \\
24 $\pm$  51 &  0.11$\pm$0.15& 0.04 $\pm$0.09&  0.00$\pm$0.04 & 5& Luck \& Heiter (\cite{luc06}) \\
\hline
\end{tabular}
\end{table*}

\section{Determination of the Li abundance}

 We used the grid of the stellar atmosphere models under the overshooting approximation from Kurucz (1993)
to compute the abundances of Li and metallicity [Fe/H].
The Li abundances in the investigated stars were obtained by fitting the observational profiles
to the synthetic spectra that were computed by the STARSP LTE spectral synthesis code, developed
by Tsymbal (\cite{tsym96}). Considering the wide temperature and metallicity ranges of the target
stars, we made every effort to compile the full list of the atomic and molecular lines close
to the \element[][7]{Li} 6707 \AA\ line (Mishenina \& Tsymbal \cite{mish97}).
Calculating the synthetic spectra, we used the values of vsini, obtained with the relation
calibrated by Queloz et al. (\cite{que98})  described above.
 The stellar parameters and the Li abundances, measured in the present study, as well as the basic
stellar characteristics are presented in Table \ref{table:tabsumm}.

Table \ref{table:3stars} shows  the example of three stars with different parameters:
the error determination for some typical models of the investigated stars,
for strong lines with EW$_{1}$ = 100 m\AA\, and for weak lines with
EW$_{2}$ = 5 m\AA\ with $\Delta \rm T_{eff}$ = +100 K (column 1); $\Delta$\logg = --0.2 (column 2);
$\Delta V_{t}$=+0.2 km/s (column 3); $\Delta EW_1$=$\pm$3 m\AA\ (column 4); and
$\Delta EW_1$=$\pm$2 m\AA\ (column 5). The total error is given in column 6.
As seen in Table \ref{table:3stars}, the total uncertainty grows for the stars with decreasing
temperatures and depends on the EW lines. It reaches 0.10--0.14 dex in the abundance determination
for strong lines (near 100 m\AA) and 0.18--0.20 dex for weak lines (near 5 m\AA).

 The comparison of the stellar parameters and the Li abundance with the results by
other authors is given in Table \ref{table:tabcomp2}.
 In Table \ref{table:comp_tak} we compared our
Li abundances with those in some papers, which have only one common star with our work.
The parameters and the Li abundance, determined
by us, are mainly in good compliance with the determinations of other authors.
 The comparison was made for the precise determinations of
the Li content, but not for the estimations of the upper limits of the Li abundance. As  marked  in the study by Lubin et al.( \cite{lub10}),  which
collected the determinations of the Li abundance, made in thirty studies
by different authors and for various objects, for log A(Li)$>$1.5 there is
a relatively small dispersion in the Li abundance detections among various measurement
programmes, while  the scattering increases for lower log A(Li). The difference in log A(Li) is about 0.5--1 dex.

%Table 7
\begin{table*}
\caption[]{The comparison of the parameters and the Li abundance for some stars}
\label{table:comp_tak}
\begin{tabular}{rlccl}
\hline
\hline
     HD  &   \Teff& \logg &    log A(Li)&                    \\
\hline
17925    &   5225 &   4.56&       2.68  &    this work         \\
         &   5235 &   4.66&       2.82  &  Takeda et al. (\cite{tak05})\\
         &   5012 &   4.60&       2.45  &  Christian et al. (\cite{chris05})\\

30495    &   5820 &   4.40&       2.35  &    this work         \\
         &   5880 &   4.67&       2.45  &  Israelian et al. (\cite{isr04})\\

190406   &   5905 &   4.30&       2.28  &    this work         \\
         &   5797 &   4.38&       2.26  &  Chen \& Zhao (\cite{chen06})\\

219623   &   5949 &   4.20&       2.60  &      this work       \\
         &   6103 &   4.18&       2.76  & Takeda et al. (\cite{tak05})\\
         &   6130 &   4.21&       2.68  & Lambert et al. (\cite{lam91})\\
\hline
\end{tabular}
\end{table*}

\section{Lithium,  stellar parameters, and activity}

  Our target stars have been selected as the stars belonging to the
  lower part of the MS upon
  photometric criterion (\Mv--(B--V)), where  \Mv = V + 5 + 2.5 log$\pi$ (M08).
  Figure \ref{LLT} shows the positions of the investigated stars at the luminosity
  log L/L$_{\sun}$ against the effective
  temperature log \Teff\ diagram where log L/L$_{\sun}$ = 0.4(4.75-- \Mbol),
  and \Mbol = \Mv + BC (the bolometric corrections BC were taken from
   Flower \cite{flo96}) and the evolutionary tracks
  by Girardi et al. \cite{gir00}.
  Those stars are  the FGK spectral type ones, and as can be seen from Fig. \ref{LLT}, they are almost evenly distributed over
the effective temperature
and have masses of about the solar one (from 0.7 to 1.1 M$_{\sun}$). They have the average value of
gravity $<$\logg$>$ equal to 4.45 $\pm$0.16
and metallicity $<$\FeH $>$ = --0.01 $\pm$0.17.
The rotational velocities of the investigated stars generally do not exceed 6 km/sec (Fig.3, upper panel). The X-ray flux distribution has just one peak and the average
value equals --4.58 $\pm$0.58; however, the distribution of the intensities of the H and K \ion{Ca}{ii} emission lines in the chromospheres has two peaks
that are separated by value \R = --4.75 (Fig.3, bottom panel), which correspond  to the values adopted by us (Section 3).
  Comparison of this histogram with similar distributions for active
  late-type stars (see, for instance, Katsova \& Livshits 2011) shows that our set of stars is characterized
by relatively moref stars with the higher activity.

The Li is detected in 43 stars among 69 stars with a high level of
the chromospheric activity, while it is detected in ten stars among 31 dwarfs with weak level of
the solar-type activity, i.e. in about  62\%  and 31\% , respectively.
We have confirmed our previous result  (\cite{mish08})  that  the frequency  of stars with the  observed Li  abundance is higher in active stars.

\subsection {Lithium abundance behaviour in relation to \Teff\ and metallicity}

 The obtained Li abundances $\log$ A(Li) are presented as a function of the effective
temperature \Teff\ in Fig. \ref{liTeff}.
Our set of stars showed a typical decrease of the Li abundance with
decreasing temperature as predicted by the theoretical calculations
(e.g. D'Antona \& Mazzitelli \cite{dan97}), but with a rather wide spread.
The active stars have higher values for the Li abundance
than the usual dwarfs at any given temperature.

The excess of Li may be explained,
in particular, by the production of additional Li in the flares. To refine that assumption, it
is necessary to study the isotopic ratio  $^{6}$Li/$^{7}$Li, which requires spectra with
resolution better than 100 000 and a signal-to-noise ratio greater than 400. That is beyond the scope
of the present study.

\begin{figure}
\includegraphics[width=8.8cm]{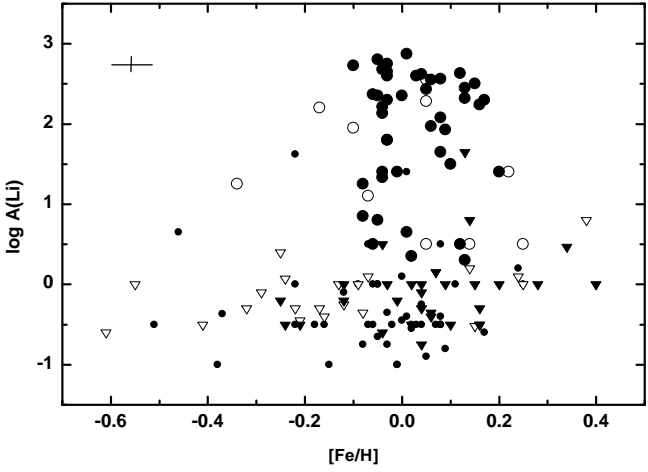}
\caption{The Li abundances vs. \FeH,
the notation is the same as in Fig.\ref{liTeff}.}
\label{lifeh}
\end{figure}

This spread is unlikely owing to the difference in
metallicity, because Fig. \ref{lifeh} with log A(Li) vs. \FeH\ shows evidence of the fairly compact
distribution of metallicity around the solar value for active stars.

%fig5

Then, we compared the Li abundance
behaviour with ``the middle course of the Li trend" for several OC's of different ages
 (in Fig.3 the trends are given schematically).
 We can see in Fig.3, the Li abundance
spread is due to different ages of stars. Younger stars have higher Li abundance, but, at the
same time, they are actually active stars.

\subsection{Relationship between the Li abundance, rotation, and the chromospheric activity index}

Below we consider the dependencies of the Li abundance
on other parameters, including the activity indices, for the stars of
three different temperature ranges: the first group with  6000$>$\Teff $>$5700,  the second group
with  5700$>$\Teff $>$5200, and  the third group with \Teff $<$5200 K.

  %fig6
  \begin{figure}
\includegraphics[width=8.8cm]{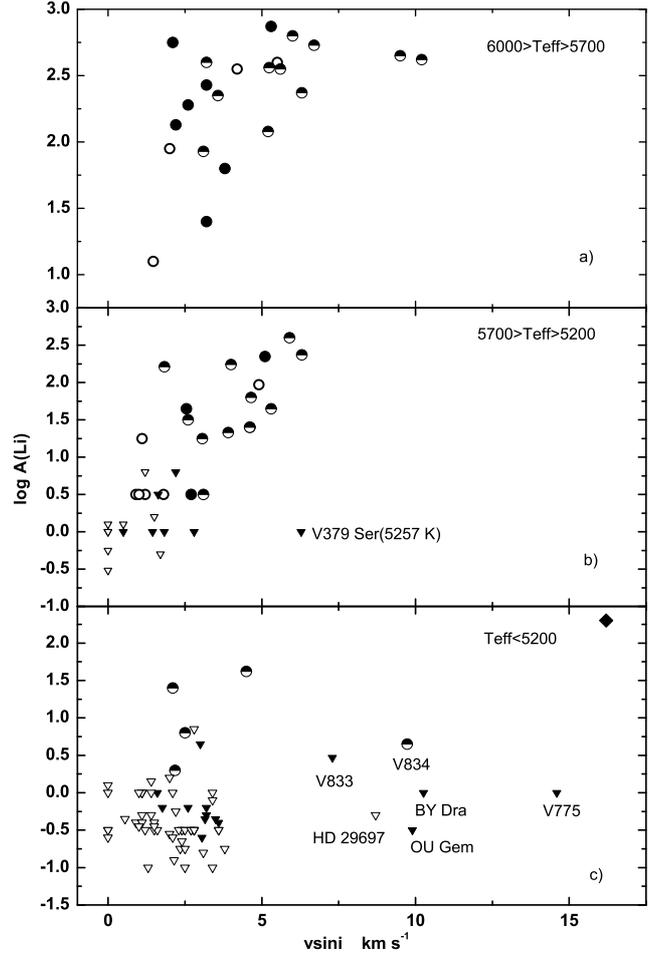}
\caption{
Dependence of the Li abundance on \vsini}
\label{Livsini}
\end{figure}

%fig7
\begin{figure}
 \includegraphics[width=8.8cm]{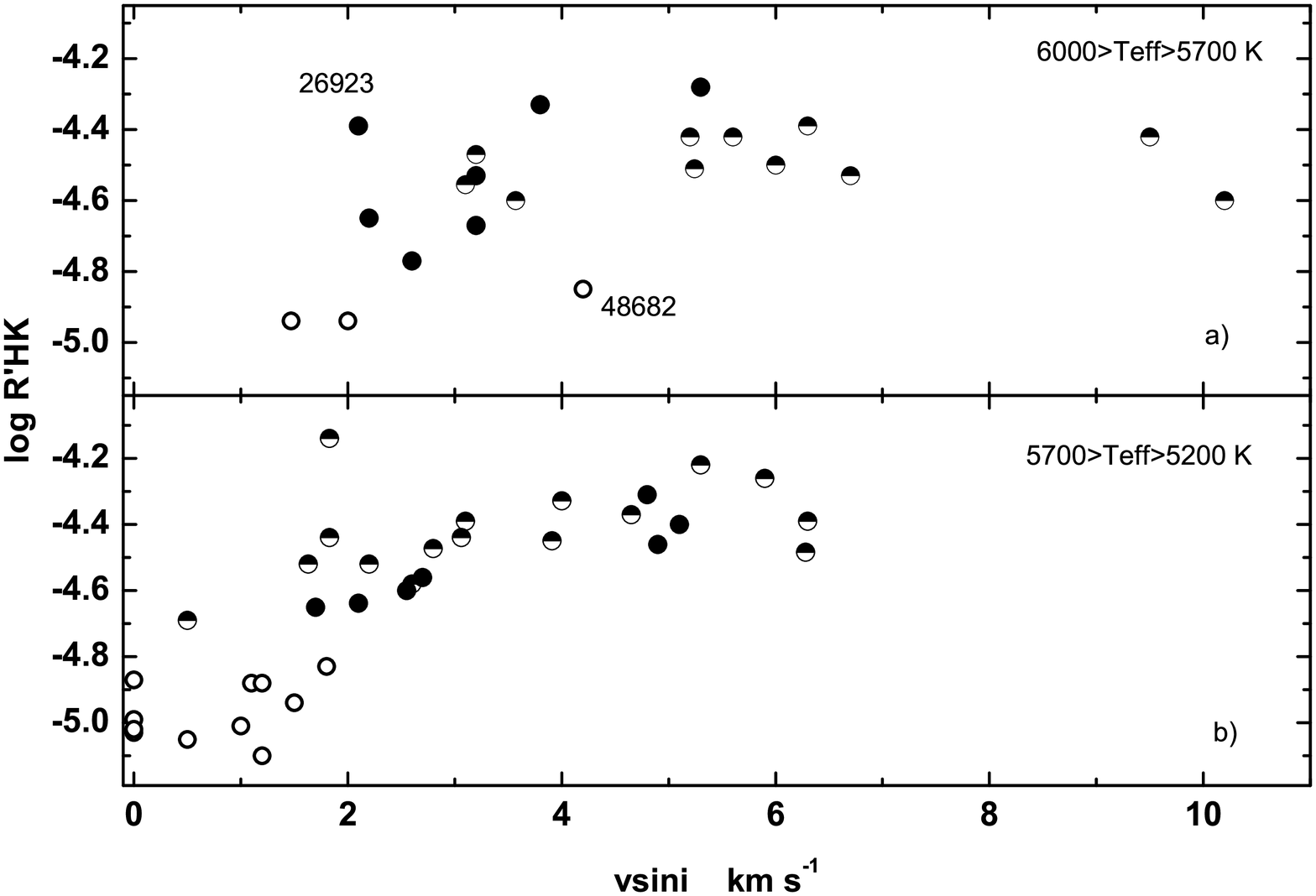}
\caption{Dependence of the \R\ on \vsini}
\label{rhkvsini}
\end{figure}

%fig8
\begin{figure}
\includegraphics[width=8.8cm]{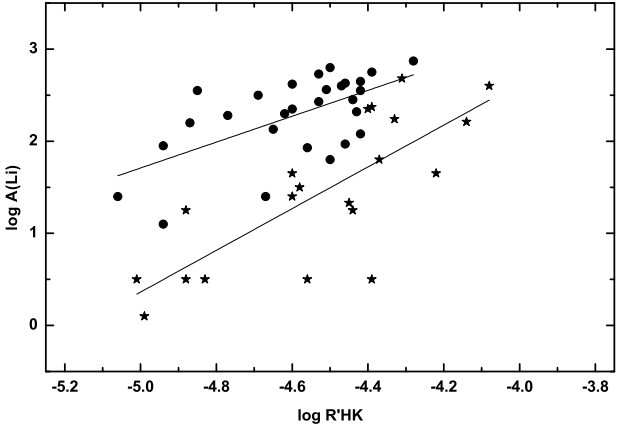}
\caption{
Dependence of the Li abundance on \R\ for the stars  with 6000$>$\Teff $>$5700 (full circles) and
  with \Teff\ from 5700 to 5200 K (asterisks)}
\label{LiTeff2}
\end{figure}

Those dependencies are quite clearly expressed for the first two groups of stars: both the Li abundance and the chromospheric activity grow
 for the faster rotators. The values of log A(Li) are
higher for the solar-type stars than for the stars cooler than the Sun (i.e. the late-type G stars). The spread of these parameters in
 Figs. 6a, b and 7 a, b is somewhat greater for the first group of stars than for the second one.

Those results can be seen more evidently  we compare the Li abundance with the chromospheric activity (Fig. 8). It is  also clear
 that the log A(Li)
 values are greater for the stars in the first group than in the second one.
The correlation coefficient is 0.64 for the solar-type stars, and 0.77 for the stars cooler than the Sun.
The spread of values of log A(Li) is less for the stars of the second group

For a few stars
with \Teff\ lower than 5200 K, a detectable value of the Li abundance was obtained.
Those stars  (see Fig. 6c) demonstrate
the absence of the correlation practically between log A(Li) and \vsini\, and there are some outliers with
high velocities and low Li abundances. Those stars, except HD 29697, are known stars of
the BY Dra type (namely, V833 Tau, V834 Tau, BY Dra, OU Gem, and V775 Her) binaries, stars with
low temperatures and very weak lines of Li in the spectra of the main component.

The lower abundances of Li for the later G stars are associated with the fact that the convection zones of those stars become deeper
 than that on the Sun.
The decrease in the Li abundance must be observed
at \Teff $<$ 5800 K where the bottom of
the surface convection zone reached the regions where Li, Be, and B burn (e.g. Michaud et al. \cite{mich04}) and
the strong convection destroys Li at \Teff $<$ 5400 K   (e.g. Iben \cite{iben65},
D'Antona \& Mazzitelli \cite{dan84} etc).

Good correlations between log A(Li) and \R\ with \vsini\ for G-type stars are evidence that the rotation is a
 key factor in the physical processes that are responsible for the levels of the Li abundance and the chromospheric activity.
Some distinctions of those dependences are due to the different role of the rotation rate in both processes.
The Li abundance depends on the axial rotation rate indirectly through the age, while the dynamo process is a direct consequence
of interaction between the rotation and the turbulent convection.

  \subsection{How can studying lithium help with understanding
  the general problems of stellar activity?}

  Up to now  stellar activity has been studied for the young objects with activity close to  saturation and less active
stars observed during execution of Exoplanet Search Programmes (for example, Wright et al. \cite{wrig04} and Mart\'{i}nez-Arn\'{a}iz et al. \cite{mart11}).
Previously, we  studied the relationship between the coronal and chromospheric activity (Katsova \& Livshits \cite{katliv06},
 \cite{katliv11}). Here we also consider  a set of stars that contains objects with a detectable Li line and with
 moderate activity levels.   Data on the chromospheric activity indices are adopted from the papers cited in  Section 3. The soft X-ray radiation
of all those stars was adopted from the ROSAT data (Schmitt \& Liefke \cite{schmitt04}, Huensch et al \cite{huen99}) and the XMM-Newton observations
(Poppenhaeger et al. \cite{Poppen10}, \cite{Poppen11}).
The data for more than 50 stars  with detected Li abundances  allows us to fill up a gap between low- and saturated activity on the "chromosphere-corona"
diagram  constructed by Katsova \& Livshits  (\cite{katliv11}) for 172 stars including the Sun.
In this way, the whole set of stars includes stars with the low- and high activity levels.

The results are presented in Fig. 9.
  Newly added stars with detected Li hardly deviate from the straight line by Mamajek \& Hillenbrand (\cite{mam08}) corresponding
to the linear regression between log \R\ and log $L_X/L_{bol}$. Most stars of that group are hotter than the Sun. Since the activity level is associated
with the age, displacement along this line describes an evolution in the activity of such stars. The activity of all late-type stars
 evolves quickly during
the first 1--2 Gyr, then their paths diverge and the stars cooler than the Sun displace below this line. Only several stars with detected Li are among
these objects, while the others, revealed earlier on the same kind of  diagram by Katsova \& Livshits (\cite{katliv11}) are less active.

Physically, the change in
the relationship between the chromospheric and coronal indices reflects the change in the properties of the activity. Probably,
that  can be due to various
depths of the convection zone in those stars and the change in conditions for the dynamo mechanism. A broad spread in estimations of the ages of
late G and K stars from their activity levels and from the Li abundance is due to that circumstance. We believe that Li  may be  a better indicator of the age
than the activity.
     We point out  that the stars, classified in the HK Project as stars with ``Excellent" cycles, are located on the diagram outside
the main group of the
F and earlier G-stars because the formation processes  of the stellar cycles can be related to the above problems.

%fig9
\begin{figure}
\includegraphics[width=8.8cm]{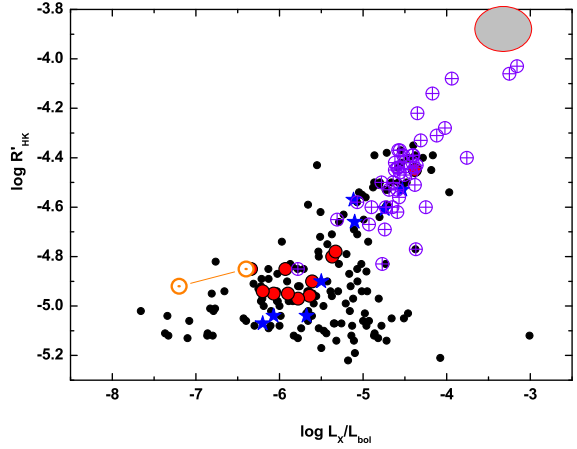}
\caption{Diagram of the chromospheric and coronal activity for the late-type stars. The stars of the basic data set are marked
as dots. The corresponding references are given above, or the summary table see Katsova \& Livshits (2011). The investigated
stars with measured Li abundance and with direct measurements of the \R-indices (from the studies mentioned above)
are marked as crosses inside the circles (violet). According to the type of a cycle, the stars of group "Excellent" are marked as red circles,
the stars of "Good" group are indicated as blue asterisks; the Sun at
the maximum and at the minimum is denoted by its own sign
and connected by the direct line.
}
\label{rhkX}
\end{figure}

 \section{Conclusion}

   The physical parameters, [Fe/H], and the Li abundances were determined homogeneously for a sample
  of 24 FGK dwarfs, and the behaviour of the Li abundance was analysed
  for 150 stars. For our sample of stars, including many BY Dra-type objects, the correlation
  between the chromospheric activity, measured by the \R-index and rotation \vsini\, and
  between the Li abundance, rotation, and the \R-index  were considered  for three
  groups of the stars with \Teff\ from 6000 to 5700 K, \Teff\ from 5700 to 5200 K
  and \Teff$<$5200 K.  The high Li abundance was detected for a small number
   of the stars with \Teff$<$5600 K.
  We can make two main conclusions.

  \begin{enumerate}
  \item The correlation Li-activity is evident only in a restricted temperature range (\Teff\ from 5700 to 5200 K).
  \item  The Li abundance spread seems to be present in a group of  low chromospheric activity stars
   that also show a broad spread in the chromospheric vs. coronal activity.

   \end{enumerate}

\begin{acknowledgements}
The authors thank the anonymous referee and Dr. Piercarlo Bonifacio, as referee, for the careful reading of the manuscript
and the important remarks that enabled the work to be improved. TM thanks the Laboratoire d'Astrophysique de Bordeaux
for their kind hospitality during the course of the present project. This study was conducted using
the SIMBAD database, operated at the CDS, Strasbourg, France. It is based on the data obtained from
 the ESA {\it  Hipparcos} satellite (Hipparcos catalogue). The present work was supported
by the Swiss National Science Foundation, project SCOPES No. IZ73Z0-128180/1. MK and ML are
grateful for the financial support within the framework of RFBR grant 12-02-00884, and RSS grant 2374.2012.2.
\end{acknowledgements}

  \section{Appendix}
HD 139813 has both high X-ray emission (\Rx=--3.75) and \R=--4.40 (Wright et al. \cite{wrig04})
that indicates it is an active star. Its SIMBAD object type is a star in a double system with high
proper motion and IR source.

HD 185414 has
\R=--4.83 according to
Baliunas et al. (\cite{bal95}). That star is in Takeda et al. (\cite{tak10}) where its measurements
indicate moderate activity (HIP 96396, r$_{0}$(8542) = 0.218).

HD 208038
is in Wright et al. (\cite{wrig04}),
but its colour is outside the boundaries for the \R\ determination. Its GrandS value (0.533)
indicates some \ion{Ca}{ii} emission, so, it can be considered as an active one.

We noted three stars with \R\ in the range --4.55 to --4.70 in Wright et al.( \cite{wrig04}), but with
no other indication of the activity (HD 4635, HD 105631, HD 184385).

 \end{document}